\documentclass[hyper]{JHEP} 

\usepackage{epsfig}  
\title{Exact Solutions in Open Bosonic String Field Theory
and Marginal Deformation in CFT}
\author{ J. Kluso\v{n}
\footnote{On leave from Masaryk University, Brno}\\
Institute of Theoretical Physics, University of Stockholm, SCFAB\\
SE- 106 91 Stockholm, Sweden \\
and \\
Institutionen f\"or teoretisk fysik\\
BOX 803, SE- 751 08 
Uppsala, Sweden \\
E-mail: \email{josef.kluson@teorfys.uu.se}}   \preprint{\hepth{0209255}}  			  	
 \abstract{In this paper we continue our study
          of the exact solution in open bosonic string field theory.
We present new  solution in the string field theory
 defined on  the background corresponding to the
boundary conformal field theory  describing
D25-brane. Then we will 
study the fluctuation modes around  this solution
and we  determine their basic properties
 from the linearised  equation
of motion of the string field theory defined
above the classical solution.}


\def\ket #1{\left|#1\right>}

\begin{document}
\section{Introduction}\label{first}
The construction of time dependent classical solutions
that describe the decay of an unstable D-brane
as the tachyon rolls down toward the minimum of
the potential has been given in series of seminal
papers by A. Sen \cite{Sen:2002nu,Sen:2002in,Sen:2002an,
Sen:2002vv,Mukhopadhyay:2002en}.
The study of time dependent solutions representing of
tachyons in $p$-adic string  theory  was performed
in \cite{Moeller:2002vx} and from different point of
views in \cite{Hashimoto:2002xt,Sugimoto:2002fp,
Minahan:2002if,Ohta:2002ac,Mehen:2002xr,
Okuda:2002yd}. Application of
the tachyon matter in cosmology has been discussed in
\cite{Gibbons:2002md,Fairbairn:2002yp,Mukohyama:2002cn,
Mukohyama:2002vq,Feinstein:2002aj,Padmanabhan:2002cp,
Frolov:2002rr,Choudhury:2002xu,Li:2002nk,
Shiu:2002qe,Padmanabhan:2002sh,
Kofman:2002rh,
Benaoum:2002zs,Sami:2002zy,Sami:2002fs,
Shiu:2002xp,Piao:2002vf,Li:2002et,Cline:2002it,
Felder:2002sv,Wang:2002sz,Bento:2002np,Buchel:2002tj,
Dvali:1998pa}. In \cite{Sen:2002nu,Sen:2002in,Sen:2002an,
Sen:2002vv} the rolling tachyon solution was
mainly discussed from the point of view
of the world-sheet boundary conformal field theory
($BCFT$). More precisely, this solution was constructed
by perturbing $BCFT$ describing the original 
D-brane by an exactly marginal perturbation. Since the
Wick rotated version is exactly solvable 
\cite{Callan:1993mw,Callan:1994ub,Polchinski:my,
Recknagel:1998ih,Gaberdiel:2001zq}
we can obtain lot of exact results considering the
rolling of the tachyon to its vacuum 
value. More precisely,
the strength $\lambda$ of the perturbation labels the
initial value of the tachyon $T$. The corresponding boundary
state gives information about the time evolution 
of various closed string sources, for recent discussion,
see \cite{Mukhopadhyay:2002en,Okuda:2002yd}. In particular,
it was shown that at $\lambda=1/2$ which corresponds to
placing the tachyon at the minimum of the potential,
the whole boundary state  vanishes at all time as
conjectured in 
\footnote{For review of the  tachyon condensation
in string field theory and
extensive list of references, see
\cite{Ohmori:2001am,
Arefeva:2001ps,DeSmet:2001af}. 
For very nice reviews of
boundary state approach to D-brane dynamics, see
\cite{DiVecchia:1999rh,DiVecchia:1999fx,
Gaberdiel:2000jr,Lerda:1999um,Sen:1999mg}}
\cite{Sen:1998ii,Sen:1998tt,Sen:1998ex,
Sen:1999mh,Sen:1999xm,Sen:1999nx}. 
It can be also shown that the rolling tachyon solution
has deep connection with spacelike branes
\cite{Gutperle:2002ai,Roy:2002ik,Kruczenski:2002ap,
Chen:2002yq}. In fact, it was stressed
recently \cite{Okuda:2002yd,Strominger:2002pc}
 that this picture is 
greatly modified for nonzero string coupling
constant $g_s\neq 0$.

In this paper we will try to present the modest 
contribution to this very reach and fascinating
subject. We will continue our study of the exact
solutions of open bosonic string field theory (SFT)
and their possible relations to  the time dependent 
solutions 
\footnote{Alternative version of the exact solution
of the open bosonic string field theory was
given recently in \cite{Kluson:2002te}.}.  
In some sense this paper can be regarded as
an application of  the
general analysis performed in
\cite{Kluson:2002kk}, for similar
 approach of the
searching of the exact solutions in SFT, see
\cite{Takahashi:2001pp,Takahashi:2002ez,
Kluson:2002kk,Kluson:2002ex,Kluson:2002hr,Kluson:2002gu}.
More precisely, we 
start with the open bosonic
string field theory defined on the background
$BCFT$ corresponding
to space-time filling D25-brane. 
Then we find such a  solution of
the SFT equation of motion that when we expand 
string field around this solution the 
 new BRST operator will have form that we could expect
for the string field theory defined around the
new conformal field theory background $BCFT''$ that
is related to  the original one by marginal deformation
in the original $BCFT$. It is important to stress
that we do not claim that the SFT for fluctuation modes
is  SFT formulated
around the background $BCFT''$. In particular, the
fluctuation modes  still belong to the Hilbert space 
of the original $BCFT$ and the correlation
functions are also calculated using the original
$BCFT$.
 In order to really show
that the SFT for fluctuation modes
 corresponds to the SFT formulated
around the new $BCFT''$ we should present such a 
careful analysis as was done in
\cite{Sen:1990hh,Sen:1990na,Sen:1992pw,
Sen:1993mh,Sen:1994kx} in case of closed string field theory.
In fact this approach is currently under study
and we hope to return to this question  in future. 

In order to study the fluctuation spectrum around
the classical solution we will proceed in different way.
Using the form of the shifted BRST operator we will 
propose such a form of the string  field  fluctuation
modes around
the classical solution that will allow straightforward
analysis of the linearised string field theory equation
of motion which these modes have to obey.
According to this analysis  we will present arguments
 that these modes  
 correspond to the states 
of the $BCFT''$  with the marginal
perturbation inserted on the boundary that
was studied in
\cite{Callan:1993mw,Callan:1994ub,Polchinski:my,
Recknagel:1998ih,Gaberdiel:2001zq} (See also
\cite{Sen:2002nu,Sen:2002in,Sen:2002an,
Sen:2002vv,Mukhopadhyay:2002en} for recent
application of this $BCFT''$  in the case of the rolling 
tachyon solution.).

However we must mention one remarkable fact.
As we will see explicitly in the next section,
 our solution has pure gauge form 
which could then  suggest that our approach 
 does not lead to any new physical situation.
But  this conclusion seems to be with contradiction
with our explicit calculations
that will show that the original  BRST operator is
really deformed by our solutions and has exactly the same form
as we could expect from deformed $BCFT$ theories studied in
\cite{Callan:1993mw,Callan:1994ub,Polchinski:my,
Recknagel:1998ih,Gaberdiel:2001zq}.
 In order to resolve this 
puzzle we will argue that there is a difference between
large gauge transformations that really reflect the
symmetry of the gauge theory, which in our  case is
string field theory,  and under which the string field theory
theory action is not generally invariant and under 
small gauge transformations that leave the action invariant and
that really does not give new physical configurations.

This paper is organized as follows. In the next section
we will calculate an exact solution of the SFT
equation of motion in the spirit of our
general analysis given in \cite{Kluson:2002kk}. 
Then we perform an expansion in the SFT action around
this solution and we will study the fluctuation
modes around the new solution that will obey
the linearised equation of motion.
We will show that the spectrum of the modes
is the same as
 in  \cite{Sen:2002nu,Sen:2002in,Sen:2002an,
Sen:2002vv,Mukhopadhyay:2002en} in the case of the
rolling tachyon solution or as in 
\cite{Callan:1993mw,Callan:1994ub,Polchinski:my,
Recknagel:1998ih,Gaberdiel:2001zq} in case of the
spatial dependent solution. 
In the conclusion we will outline our results and we 
will discuss other problems that deserve further study.


\section{Marginal solution}
In this section we will 
find exact solution
of the SFT equation of motion that can be
directly related to the  rolling tachyon solution, or more 
generally to $BCFT$ studied in
\cite{Callan:1993mw,Callan:1994ub,Polchinski:my,
Recknagel:1998ih,Gaberdiel:2001zq}.

To begin with, we firstly review basic facts about
bosonic string field theory, following
mainly \cite{Ohmori:2001am,
Arefeva:2001ps,DeSmet:2001af}. Gauge invariant
string field theory is described with the
full Hilbert space of the first quantized open
string including $b,c $ ghost fields subject
to the condition that the states must
carry ghost number one, where $b$ has ghost
number $-1$, $c$ has ghost number $1$ and
$SL(2,C)$ invariant vacuum $\ket{0}$
carries ghost number $0$. We denote
$\mathcal{H}$ the subspace of the full Hilbert
space carrying ghost number $1$.  Any state
in $\mathcal{H}$ will be denoted as $\ket{\Phi}$
and corresponding vertex operator $\Phi$
is the  vertex operator that  creates state $\ket{\Phi}$
out of the vacuum state $\ket{0}$
\begin{equation}
\ket{\Phi}=\Phi \ket{0} \ .
\end{equation}
Since we are dealing with open string theory,
the vertex operators should be put on the boundary
of the world-sheet.   The string field theory action
is given
\cite{Witten:1985cc}
\begin{equation}\label{actionCFT}
S=-\frac{1}{g_0^2}\left(
\frac{1}{2\alpha'}\left<I\circ \Phi (0) Q \Phi (0)\right>+
\frac{1}{3}\left<f_1\circ \Phi(0) f_2\circ \Phi(0)
f_3\circ \Phi(0)\right>\right) \ ,
\end{equation}
where $g_0$ is open string coupling constant, $Q$
is BRST operator and $<>$ denotes correlation function
in the combined matter ghost conformal field theory.
$I,f_1, f_2, f_3$ are conformal mapping  exact
form of which is reviewed in \cite{Ohmori:2001am} and $f_i\circ\Phi(0)$ denotes
the conformal transformation of $\Phi(0)$ by $f_i$. For
example, for $\Phi$ a primary field of dimension $h$,
then $f_i\circ \Phi(0)=(f'_i(0))^h\Phi(f_i(0))$. 
In the abstract language pioneered in 
\cite{Witten:1985cc}
the  open string field theory action (\ref{actionCFT})
is
\begin{equation}\label{actionW}
S=-\frac{1}{g_0^2}\left(
\frac{1}{2\alpha'}\int \Phi \star Q\Phi
+\frac{1}{3}\int \Phi\star\Phi\star\Phi\right) \
\end{equation}
from which we immediately get an equation of motion
\begin{equation}\label{eq}
\frac{1}{\alpha'}Q\Phi_0+\Phi_0\star \Phi_0=0  \ .
\end{equation}
It is easy to see that the string field in the form
\begin{equation}\label{Proll}
\Phi_0=e^{-K_L(\mathcal{I})}\star \frac{1}
{\alpha'}Q( e^{K_L(\mathcal{I})})
\end{equation}
is solution of (\ref{eq}) for any ghost number zero
operator $K_L$ acting on the string field theory algebra $\star$
 identity   $\mathcal{I}$ which is ghost number zero
field that obeys
\cite{Horowitz:dt} 
\begin{equation}
\mathcal{I}\star X=X\star \mathcal{I}=X \ ,
\end{equation}
for any string field $X$
\footnote{For recent study of the identity element
$\mathcal{I}$, see
\cite{Ellwood:2001ig,Matsuo:2001yb,Kishimoto:2001ac,
Schnabl:2002gg,Kishimoto:2001de}.}.
The operator $K$ in (\ref{Proll}) can be written
as 
\begin{equation}\label{K}
K=\frac{A}{2\pi i}\int_C
dz k(z)+\frac{B}{2\pi i}\int_C 
d\overline{z}\tilde{k}(\overline{z}) \ ,
\end{equation}
where  $k(z), \tilde{k}(\overline{z})$ are holomorphic and 
anti holomorphic fields of conformal
dimensions  $(1,0), (0,1)$ respectively that are
 defined in the
upper-half  plane $\mathrm{Im} \ z \geq 0$.
The field $k(z)$  transforms under  conformal transformation
$z\rightarrow f(z)$
as
\begin{equation}
U_f k(z)U_f^{-1}=\frac{df(z)}{dz}k(f(z))\equiv
f'(z)k(f(z)) \ ,
\end{equation}
with the  same expression for anti holomorphic field 
$\tilde{k}(\overline{z})$.
In general, the integral in (\ref{K}) depends on the 
integration
contour $C$ which we choose as a semicircle in the upper 
half-plane $C:\left\{ z; z=-e^{-i\sigma+\tau} \ , \sigma
\in (0,\pi /2)\right\}$ and we also
choose $\tau=0$. The fact that $\sigma \in (0,\pi/2)$ means
that we perform integration over the left side of the 
string so that we label corresponding operator with the 
subscript $L$. Operators without any subscript correspond to
the integration of any local density over the whole string so that
$\sigma \in (0,\pi)$ and finally operator corresponding
to the integration over right side of the string will be labeled
by  the subscript $R$ with $\sigma \in (\pi/2,0)$.

Now we will study the transformation properties of
$K$ (\ref{K}) under general conformal transformations.
Generally $K$ is not invariant under 
conformal transformation under which it transforms
as
\begin{eqnarray}\label{confK}
U_fKU_f^{-1}=
\frac{A}{2\pi i}\int_C dw  f'(w) k(f(w))
+\frac{B}{2\pi i}\int_C d\overline{w}
\overline{f}'
(\overline{w})\tilde{k}(\overline{f}
(\overline{w}))
= \nonumber \\
=\frac{A}{2\pi i}\int_{f(C)} df 
 k(f)+
\frac{B}{2\pi i}\int_{f(C)} d\overline{f}
\tilde{k}(\overline{f})\neq K
 \ , \nonumber \\
\end{eqnarray}
which is not  equal to $K$ since, as we have
 argued above, the integration in $K$ explicitly depends
of the integration contour  $C$. 
However we can easily prove that
for the $BCFT$ with  standard
 Neumann boundary conditions  
\begin{equation}
k(z)=\tilde{k}(\overline{z}) \ ,
\mathrm{Im} \  z=0  
\end{equation}
the operator $K$ with $A=-B$ 
will be invariant under conformal transformations
 since then the definition
of $K$ does not depend on the integration contour $C$.
To see this let us consider  following
integral over closed curve $C$ in the upper half-plane 
\begin{equation}
F(C)=\frac{1}{2\pi i}\oint_{C}dz k(z)-
\frac{1}{2\pi i}\oint_{C}d\overline{z}\tilde{k}(
\overline{z}) \ .
\end{equation}
This integral  
 is equal to zero 
thanks to the holomorphicity and antiholomorphicity of
$k, \tilde{k}$ respectively. 
Let us consider  $C$ as
\begin{eqnarray}
C=C_1+C_{r1}+C_{r2}+C_2' \ ,\nonumber \\
C_1=\left\{z; z=-r_1e^{-i\sigma} \ , \sigma\in (0,\pi) 
\right\} \ ,\nonumber \\
C_{r1}=\left\{  z\in (r_1,r_2) \ ,
\mathrm{Im} \  z=0 \ , r_1>r_2 \right\} \ ,
\nonumber \\
C_2'=\left\{z; z=-r_2e^{-i\sigma}\  , \sigma\in (\pi,0) 
\right\} \ ,\nonumber \\
C_{r2}=\left\{z \in (r_2,r_1) \ , \mathrm{Im} \  
z=0 \ \right\} \ .
\nonumber \\
\end{eqnarray}
Then it is easy to see that  the
integrals  of 
$k,\tilde{k}$ performed over
 $C_{r1}, C_{r2}$ cancel each
other  since $z=\overline{z} \ , \ k(z)=
\tilde{k}(\overline{z})$. 
Consequently we get
\begin{equation}
F(C)=0=
K(C_1)+K(C_2')=K(C_1)
-K(C_2) \Rightarrow K(C_1)=K(C_2)
\ , C_2'=-C_1
\end{equation}
so that $K$  does not depend on the
integration contour $C$  in the upper-half  plane. 
For such an operator (\ref{confK}) gives 
\begin{equation}
U_fKU_f^{-1}=K \ .
\end{equation}
Now we prove that the operator $K$ defined in
(\ref{K}) obeys 
\cite{Horowitz:dt}
\begin{equation}\label{derK}
K_L(A)\star B=-A\star K_R(B)
\end{equation}
for any string fields $A \ ,B$. We  follow
 mainly 
\cite{Takahashi:2002ez} which is based on
fundamental papers
\cite{LeClair:1988sp,LeClair:1988sj}.
The $N$-string vertex of a midpoint interaction
is defined by gluing the boundaries 
$|w_i|=1 \ , (i=1,\dots, N) $ of $N$ unit half disks
in the upper-half plane $\mathrm{Im} \ z \geq 0$
 with the identifications
\begin{equation}\label{id}
w_iw_{i+1}=-1 \ ,
\mathrm{for} \  |w_i|=1 \ , \mathrm{Re} \  w_i \leq 1 \ .
\end{equation}
Under this identification the primary 
field of conformal
dimension $h$ transforms as
\begin{equation}
U_f\phi(w_{i+1})U_f^{-1}=\left
(\frac{df(w_{i+1})}{dw_{i+1}}\right)^h
\phi(f(w_{i+1})) \ ,  f(w_{i+1})=w_i \ 
\end{equation}
so that for $\phi(w)$ that is  primary field of conformal
 dimension $1$ we
get 
\begin{equation}
U_{f}dw_{i+1}\phi(w_{i+1})U_f^{-1}
=dw_{i+1}\left(\frac{d w_i}{dw_{i+1}} 
\right)\phi(w_i)=
dw_i \phi(w_i)
\end{equation}
and similar for complex conjugate $\overline{w}_i,
\overline{w}_{i+1}$. 
Using this result and the 
fact that under identification (\ref{id}) 
we change the range of integration from $\sigma\in
(0,\pi/2)$ to $(\pi,\pi/2)$ we  
immediately get (\ref{derK}).   
For the spatial case $A=B=
\mathcal{I}$  (\ref{derK}) gives
\begin{equation}\label{derKI}
K_L(\mathcal{I})=K_L(\mathcal{I})\star
\mathcal{I}=-\mathcal{I}\star 
K_R(\mathcal{I})=
-K_R(\mathcal{I}) \ .
\end{equation}
It is important to stress that we have 
got this result (\ref{derKI})
 without presumption that
$K$ is the derivation of  the star algebra 
\begin{equation}
K(A\star B)=K(A)\star B+A \star K(B) \ .
\end{equation}
On the other hand  it is well known that the
BRST operator obeys 
\begin{equation}
Q(A\star B)=Q(A)\star B+(-1)^{|A|}A
\star Q(B) 
\end{equation}
which for $A=\mathcal{I}$ implies $Q(\mathcal{I})
=Q_L(\mathcal{I})+Q_R(\mathcal{I})=0 $
\cite{Horowitz:dt}.

In the following we restrict ourselves to the
 case $B=0$ in (\ref{K})
 since we will not demand
invariance of $K$ under conformal transformation. 
In fact in our further discussion
 we will consider
two particular examples 
of $k$
\footnote{In the following we will often
omit the subscripts $t,s$ in  situations, where
we will not need explicit form of $k$.}
\begin{equation}\label{KT}
K_t=\frac{A}{2\pi i}
\int_C dz k_t(z)\equiv \frac{A}{2\pi i}\int_C dz 
\exp \left(\frac{2}{\sqrt{\alpha'}}
X^0_L(z)\right)  \ ,
\end{equation}
\begin{equation}\label{Ks}
K_s=\frac{A}{2\pi i}
\int_C dz k_s(z)\equiv\frac{A}{2\pi i}\int_C dz
\exp \left(i\frac{2}{\sqrt{\alpha'}}
X^1_L(z)\right) \ .
\end{equation}
Now the  fields
$k_t(z),k_s(z)$ have  conformal dimension
 equal to $1$  as can be
seen from the following OPE 
\begin{equation}
T(z)k_{s,t}(w)=
\frac{1}{(z-w)^2}k_{s,t}(w)+\frac{1}{z-w}\partial
k_{t,s}(w) \  ,
\end{equation}
where in  our calculation we use convention from
the very nice review
\cite{Ohmori:2001am}
\begin{eqnarray}\label{conM}
T_m(z)=-\frac{1}{\alpha'}\partial_z X^{\mu}_L(z)
\partial_z X^{\nu}_L\eta_{\mu\nu} 
 \ , \nonumber \\
\tilde{T}_m(\overline{z})=
-\frac{1}{\alpha'}\partial_{\overline{z}}
 X^{\mu}_R(\overline{z})
\partial_{\overline{z}} 
X^{\nu}_R(\overline{z})\eta_{\mu\nu} 
 \ , \nonumber \\
 X^{\mu}_L(z)X^{\nu}_L(w)\sim
-\frac{\alpha'}{2}\eta^{\mu\nu}
\ln (z-w)  \ , \nonumber \\
 X^{\mu}_R(\overline{z})
X^{\nu}_R(\overline{w})\sim
-\frac{\alpha'}{2}\eta^{\mu\nu}
\ln (\overline{z}-\overline{w})  \ , 
\nonumber \\
X^{\mu}_L(z)X^{\nu}_R(\overline{w})\sim
-\frac{\alpha'}{2}\eta^{\mu\nu}
\ln (z-\overline{w})  \  
\nonumber \\
\end{eqnarray}
with the BRST operator 
\begin{eqnarray}
Q=\frac{1}{2\pi i} \int_C
dz j_B(z)-\frac{1}{2\pi i}\int_C d
\overline{z}\tilde{j}_B(\overline{z}) \ , \nonumber \\
j_B(z)=c(z)
\left[T_m(z)+\frac{1}{2}T_{gh}(z)\right] \ ,
\nonumber \\
\tilde{j}_B(\overline{z})=
\tilde{c}(\overline{z})
\left[\tilde{T}_m(
\overline{z})+
\frac{1}{2}\tilde{T}_{gh}(
\overline{z})\right] \ ,
\nonumber \\
\end{eqnarray}
where $j(z)$ is holomorphic and $\tilde{j}(\overline{z})$
 is anti holomorphic
current  and where $T_{ghost}$ is the stress energy tensor
for the ghost field. 
In what follows we will not need
to know the  explicit form of the ghost contribution.

In summary, we have shown in 
the previous part that there are
 exact solutions of the SFT equation of motion
that are based on the existence of primary fields of
conformal dimension  $1$. Using these fields we 
have constructed
operators in $BCFT$  that are invariant under 
conformal transformations. This implies, since these
operators do not contain ghost contribution, 
 that their commutator with 
the  BRST charge is nonzero. This important property
the fact that these operators obey (\ref{derK}) 
leads to the modification  of the BRST operator 
in the original $BCFT$ theory
to the new form of the BRST operator $Q'$. 
It is important to stress that this operator $Q'$ 
 is  still defined in the original
$BCFT$ theory, however the form of the  action
for fluctuation modes
suggests that the string field action formulated around
the classical solution  corresponds to the
string field theory action formulated around new
$BCFT''$ that arises from the original
$BCFT$ by marginal interaction inserted
on the boundary of the world-sheet.

To see this more explicitly,
 let us expand the string field $\Phi$ in
(\ref{actionW}) around the solution  (\ref{Proll}) as
\begin{equation}
\Phi=\Phi_0+\Psi \ .
\end{equation}
and insert it  in (\ref{actionW}).
Then we obtain an
action for the fluctuation field $\Psi$ in
the same form as  the original one
(\ref{actionW}) 
\begin{equation}\label{actionWF}
S=-\frac{1}{g_0^2}\left(
\frac{1}{2\alpha'}\int \Psi \star Q'\Psi
+\frac{1}{3}\int \Psi\star\Psi\star\Psi\right) \ , 
\end{equation}
where    the new  BRST operator
$Q'$ is introduced 
\cite{Horowitz:dt}
\begin{equation}\label{Q'}
Q'(X)=Q(X)+\Phi_0\star X-
(-1)^{|X|}X\star \Phi_0 \ .
\end{equation}
In order to obtain the new form of the BRST
operator (\ref{Q'}) we will follow the calculation
outlined in \cite{Kluson:2002kk} 
and which we
briefly review below. 
We start with the function
\begin{equation}
F(t)=\frac{1}{\alpha'}
e^{-K_L(\mathcal{I})t}\star
Q(e^{K_L(\mathcal{I})t}) \ \ ,
F(1)=\Phi_0 \ , F(0)=0 
\end{equation}
and perform Taylor expansion around the
point $t=1$
\begin{equation}
\Phi_0=F(1)=F(0)+
\sum_{n=1}^{\infty}
\frac{1}{n!}
\frac{d^nF}{d^nt}(0) \ ,
\end{equation}
where
\begin{eqnarray}
\frac{dF}{dt}=\frac{1}{\alpha'}
e^{-K_L(\mathcal{I})t}
\star [Q,K]_L(\mathcal{I})\star 
e^{K_L(\mathcal{I})t} \ , 
\nonumber \\
\frac{dF}{dt}(0)=\frac{1}{\alpha'}
[Q,K]_L(\mathcal{I})\equiv
D_L(\mathcal{I}) \ , 
\nonumber \\
\frac{d^2F}{d^2t}=
e^{-K_L(\mathcal{I})t}(-K_L(\mathcal{I})
\star D_L(\mathcal{I})+
D_L(\mathcal{I})\star K_L(\mathcal{I})
)\star e^{K_L(\mathcal{I})t} 
\ , \nonumber \\
\frac{d^2F}{d^2t}(0)=-K_L(\mathcal{I})
\star D_L(\mathcal{I})+
D_L(\mathcal{I})\star K_L(\mathcal{I})=
[K,D]_L(\mathcal{I}) \ , \nonumber \\ 
\frac{d^3F}{d^3t}(0)=
[K_L,[K_L,D_L]]  
\ , 
\dots  \ ,
\frac{d^nF}{d^nt}(0)=\overbrace{[K,[K,\dots,
[Q,K]]]_L}^{n-1}
(\mathcal{I}) 
\nonumber \\
\end{eqnarray} 
and consequently
\begin{equation}\label{phi}
\Phi_0=\frac{1}{\alpha'}
\sum_{n=1}^{\infty}
\frac{1}{n!}\overbrace{[K,[K,\dots,[Q,K]]]_L}^{n}
(\mathcal{I})\equiv 
\mathcal{D}_L(\mathcal{I})\ .
\end{equation}
From  the upper expression we
 see that we can express $\Phi_0$ 
as an action 
of the ghost number one operator
$\mathcal{D}_L$ acting on the identity field. 
Then we immediately
obtain
\begin{eqnarray}\label{Qgen}
Q'(X)=Q(X)+\mathcal{D}_L(\mathcal{I})
\star X-(-1)^{|X|}X\star \mathcal{D}_L
(\mathcal{I})= \nonumber \\
=Q(X)-\mathcal{I}\star 
\mathcal{D}_R(X)-\mathcal{D}_L(X)
\star \mathcal{I}=
Q(X)-\mathcal{D}(X) \ . 
\nonumber \\
\end{eqnarray}
In the calculation performed above it is important
to stress that if two operators $Q,K$ are derivatives
of the string field algebra
\begin{eqnarray}
Q(A\star B)=Q(A)\star B+(-1)^{|A|}A\star Q(B) \ ,
\nonumber \\
K(A\star B)=K(A)\star B+A\star K(B) \nonumber \\
\end{eqnarray}
then so is its commutator. 
To see this, let us define
$C\equiv [Q,K]$. Then we can write
\begin{eqnarray}
C(A\star B)=Q(K(A)\star B+A\star K(B))
-K(Q(A)\star B+(-1)^{|A|}A\star Q(B))=\nonumber \\
=QK(A)\star B+(-1)^{|A|}K(A)\star Q(B)
+Q(A)\star K(B)+(-1)^{|A|}A\star QK(B)-\nonumber \\
-KQ(A)\star B-Q(A)\star K(B)
-(-1)^{|A|}K(A)\star Q(B)-(-1)^{|A|}A\star KQ(B)=\nonumber \\
=[Q,K](A)\star B+(-1)^{|A|}A\star[Q,K](B)
=C(A)\star B+(-1)^{|A|}A\star C(B) \ .
\nonumber \\
\end{eqnarray}
Now  it is clear
that the BRST operator $Q'$ is completely 
determined by commutators of $K$ with
$Q$. This calculation can be easily performed 
using
\begin{eqnarray}
\left[\frac{1}{2\pi i}
\int dz c(z)T_m(z),\frac{A}{2\pi i}
\int_C dw k(w)\right]=
\nonumber \\
=\frac{A}{2\pi i}\int_{C} dw
\left[\frac{1}{2\pi i}\oint_{C'}dz c(z) T_m(z)
k(w)\right]=
\nonumber \\
=\frac{A}{2\pi i} \int_{C}dw
\left[\frac{1}{2\pi i}\oint dz
c(z)\left(\frac{1}{(z-w)^2}
k(w)+
\frac{1}{z-w}\partial_w 
k(w)\right)
\right]=\nonumber \\
=\frac{A}{2\pi i}\int_{C}
dw  \partial_w \left(
c(w)k(w)\right)=\frac{A}{2\pi i}
\left.c(w)
k(w)
\right|_{-\epsilon}^\epsilon \ .  \nonumber \\
\end{eqnarray}
In the previous expression $C'$ is a  small circle
around the point $w$ on  $C$. 
Note that the operator $K$ explicitly
depends on the integration contour which in our
case is semicircle of radius $\epsilon \ll 1$  in the upper half-plane 
$\mathrm{Im}  \ z \geq 0$. In fact, we have
defined operator $K$ by action  on the general 
string field which is functional 
of string coordinates $X(\sigma,\tau=0)$ evaluated
at time $\tau=0$ so that $C$
is a  semicircle of
 radius $\epsilon $ around the point $z=0$.
 As usual we take
the limit $\epsilon \rightarrow 0$ 
in the end of calculation.   On the other hand 
$Q$ is conserved charge  so that we can deform
the integration contour on which is defined
and then using   
OPE between $T_m$ and $k$ we  can
calculate the integral over $C'$.

We can also 
see from (\ref{conM}) that there is
nontrivial OPE between $X_L,X_R$ 
as a consequence of the boundary conditions at
 $\mathrm{Im} \  z =0$
imposed on the fields $X=X_L+X_R$. 
This suggests
that  we should calculate following term
as well
\begin{equation}
\left[-\frac{1}{2\pi i}
\int d\overline{z}
\tilde{c}(\overline{z})
\tilde{T}_m(\overline{z}),
\frac{A}{2\pi i}
\int_{C}
 dw k (w) 
 \right] \ .
\end{equation}
However it is easy to see that this 
expression is equal to zero. 
This follows from the
fact that  the OPE between $\tilde{T}(\overline{z})
k(w)$ contains terms such as $1/(\overline{z}-w)$. 
On the other hand, if $Q$ is given as an integral of
holomorphic and anti holomorphic currents 
defined in the region  $\mathrm{Im} \ z  \geq 0$ 
then the integral over $\overline{z}$ 
lies in   the lower half-plane $\mathrm{Im} \ z \leq  0$.
Then it is clear that the integral over
complex conjugate of  $C'$  does not encircle any singularity
and hence it  is equal to zero. 
The only possible dangerous terms
could arise from the  real axis however we can avoid these problems
by defining the integration contour $C$ in $K$ 
starting slightly above the real axis, say at 
$z=-e^{-i\epsilon} \ \epsilon \ll 1$ and then 
after working out the previous commutator we take 
the limit $\epsilon \rightarrow 0$.
As a result, the first term in 
(\ref{phi}) is equal to
\begin{equation}
\left[Q,K\right]=
\frac{A}{2\pi i}\left(c(\epsilon)k(\epsilon)-
c(-\epsilon)k(-\epsilon)\right) \ .
\end{equation}
Next term is equal to
\begin{equation}
\left[K,\left[Q,K\right]\right]
\end{equation}
Now it is easy to see that further commutators
in (\ref{phi}) are equal to zero thanks to
the trivial  OPE between
$k(z)k(w)$. 
Finally we obtain the resulting BRST operator
\begin{equation}\label{Qshift}
Q'=Q-\frac{A}{2\pi i}
c(\epsilon)k(\epsilon)+\frac{A}{2\pi i}
c(-\epsilon)k(-\epsilon) \ .
\end{equation}
Now we come to the interpretation of  this result. It is clear
from (\ref{Qshift}) that SFT
expanded around the classical solution is  in some sense
related to the deformed $BCFT''$ that arises
from the original one by 
 marginal deformation inserted on the boundary
of the world-sheet
\begin{equation}\label{Sbound1}
S_{boud}=-A\int dt  k(t) \ .
\end{equation}
The question is what means the claim ``in some sense
related''. First of all,  the fluctuation string field
still belongs to the Hilbert space  of the original  $BCFT$. 
The correlation functions which
define SFT action in the CFT formalism are also
evaluated in the original $BCFT$ so it is certainly
false to claim that we now have SFT formulated around
the new background $BCFT''$. 
To find relation between SFT  action for fluctuation
field  and
SFT action defined on the  $BCFT''$ theory 
we should perform the same analysis as in
very nice papers in \cite{Sen:1994kx,Sen:1993mh,
Sen:1990na,Sen:1990hh,Sen:1992pw}.
Since this analysis is rather complicated it 
deserves separate publication and
we hope to obtain some explicit results in the near future.

We must also stress that the previous example is strictly
speaking valid for rolling tachyon solution  since
only in this case the operator $K$ is real. For
that reason we will rather consider
following operators $K_{2s,t}$
\begin{eqnarray}\label{K2c}
K_{2s,t}=\frac{A}{4\pi i}\int_C dz
k^+(z)+\frac{A}{4\pi i}\int_C dz
k^-(z) \ , \nonumber \\
k^+_t(z)=\exp \left(\frac{2}{\sqrt{\alpha'}}
X^0_L(z)\right) \ , \
k^-_t(z)=
\exp \left(-\frac{2}{\sqrt{\alpha'}}
X^0_L(z)\right) \ , \nonumber \\
k^+(z)_s=\exp \left(\frac{2i}{\sqrt{\alpha'}}
X^1_L(z)\right) \ , \ 
k^-_s(z)=
\exp \left(-\frac{2i}{\sqrt{\alpha'}}
X^1_L(z)\right) \ .\nonumber \\
\end{eqnarray}
Since the operators $k^+_{s,t},
k^-_{s,t}$ have 
conformal weights $1$ we
immediately see that $K_{2s,t}$ obeys the same
rules as $K$ in (\ref{K}) so that the string
field (\ref{Proll}) with $K$ given above
is the  exact  solution of SFT equation of motion.

To obtain the form of the shifted BRST operator
we proceed as in the previous case. Now 
 the first commutator
is equal to
\begin{equation}\label{fk2}
\left[Q,K_{2s,t}\right]=
\frac{A}{4\pi i}\left(c(\epsilon)k^+_{s,t}(\epsilon)-
c(-\epsilon)k^+_{s,t}(-\epsilon)\right)+
\frac{A}{4\pi i}\left(c(\epsilon){k}^-_{s,t}(\epsilon)-
c(-\epsilon)k^-_{s,t}(-\epsilon)\right) \ .
\end{equation}
The  second term in  (\ref{phi})  is equal to 
\begin{equation}
\frac{1}{2\alpha'}
\left[K_{2s,t},[Q,K_{2s,t}]\right]  \ 
\end{equation}
which using (\ref{fk2}) gives
\begin{eqnarray}\label{sa}
\left[K_{s,t},c(\tau)\frac{A}{4\pi i}
k^+_{s,t}(\tau)\right]_{\tau=\epsilon,-\epsilon}=
\frac{A}{4\pi i}
\left[\frac{1}{4\pi i}\oint_C dw
k^-_{s,t}
(w),c(\tau)k^+_{s,t}(\tau)\right]_{\tau=\epsilon,-\epsilon}
=\nonumber \\
=\left.\frac{A}{8\pi i}
\frac{1}{2\pi i}
\oint_C dw c(\tau)\left(
\frac{1}{(w-\tau)^2}-\frac{1}{w-\tau}
k^3_{s,t}(\tau)\right)\right|_{\tau=-\epsilon,\epsilon}
=\nonumber \\
=\left.\frac{A}{8\pi i}
\frac{\partial}{\partial_{\tau}}
\left(\frac{1}{2\pi i}
\oint_C dw \frac{1}{w-\tau}\right)c(\tau)
-\frac{A}{8\pi i}
c(\tau)k^3_{s,t}(\tau)\right|_{\tau=-\epsilon,\epsilon}=
-\left.\frac{1}{8\pi i}c(\tau)
k^3_{s,t}(\tau)\right|_{\tau=-\epsilon,\epsilon} \nonumber \\
\end{eqnarray}
and
\begin{eqnarray}\label{sb}
\left[K_{s,t},\frac{A}{4\pi i}
c(\tau)k^-_{s,t}
(\tau)\right]_{\tau=-\epsilon,\epsilon}=
\left[\frac{A}{4\pi i}
\int dw k^+_{s,t}(w),
\frac{A}{4\pi i}
c(\tau)k^-_{s,t}(\tau)\right]_{\tau=-\epsilon,\epsilon}
=\nonumber \\
=\frac{A}{8\pi i}\frac{1}{2\pi i}
\int dwc(\tau)
\left.\left(
\frac{1}{(w-\tau)^2}+
\frac{1}{w-\tau}k^3_{s,t}(\tau)\right)
\right|_{\tau=-\epsilon,\epsilon}=
\left.\frac{A}{8\pi i}c(\tau)
k^3_{s,t}(\tau)\right|_{\tau=-\epsilon,\epsilon} \ . \nonumber \\
\end{eqnarray}
In the previous calculations we used
\begin{eqnarray}
k^+_{s,t}(z)k^-_{s,t}(w)=
\frac{1}{(z-w)^2}+\frac{1}{z-w}k^3_{s,t} \ ,
\nonumber \\ 
k^3_t(z)=\frac{2}{\sqrt{\alpha'}}
\partial_z X^0_L(z) \ , 
k^3_s(z)=\frac{2i}{\sqrt{\alpha'}}
\partial_z X^1_L(z)         
\ . \nonumber \\
\end{eqnarray}
From (\ref{sa}), (\ref{sb}) we immediately
see that   the second  term in (\ref{phi}) is equal 
to zero and consequently (\ref{fk2}) is  the
only nonzero term in  (\ref{phi}).  
If we choose 
\begin{equation}
A=-2\pi\lambda 
\end{equation}
we  obtain the shifted BRST operator
\begin{eqnarray}\label{Qshift2}
Q'_{s,t}=Q+\frac{i\lambda}{2}
c(\epsilon)k^+_{s,t}(\epsilon)-
\frac{\lambda}{2}c(-\epsilon)k^+_{s,t}(-\epsilon)+
\nonumber \\
+\frac{i\lambda}{2}c(\epsilon)k^-_{s,t}(\epsilon)-
\frac{\lambda}{2}
c(-\epsilon)k^-_{s,t}(-\epsilon)
\Rightarrow\nonumber \\
\Rightarrow Q'_t=Q+i\lambda c(\epsilon)\cosh \left(
\frac{2}{\sqrt{\alpha'}}X^0(\epsilon)\right)
-i\lambda c(-\epsilon)\cosh \left(
\frac{2}{\sqrt{\alpha'}}X^0(-\epsilon)\right) \ ,
\nonumber \\
Q'_s=Q+i\lambda c(\epsilon) \cos \left(
\frac{2}{\sqrt{\alpha'}}X^1(\epsilon)\right)
-i\lambda c(-\epsilon)\cos \left(
\frac{2}{\sqrt{\alpha'}}X^1(-\epsilon)\right) \ ,
\nonumber \\ 
\end{eqnarray}
where $Q_t$  corresponds to the
 $k_t^{+,-}$ in (\ref{K2c}). With the analogy
with the previous calculation  we would like to interpret
 this deformation of the BRST operator
as the result of the  marginal
interaction inserted on the boundary of 
the world-sheet
\begin{equation}\label{b1}
S_{bound}=\lambda \int d\tau
\cosh\left(\frac{2}{\sqrt{\alpha'}}X^0(\tau)\right) \ .
\end{equation}
This $BCFT$ was extensively studied in 
\cite{Sen:2002nu,Sen:2002in,Sen:2002an,
Sen:2002vv,Mukhopadhyay:2002en} and
corresponds to the rolling tachyon solution. 
Once again we must stress that this claim
should be proved in the same way as in
\cite{Sen:1994kx,Sen:1993mh,Sen:1990na,
Sen:1990hh,Sen:1992pw}. At this place we have
only shifted BRST operator in the Hilbert
space of the original $BCFT$ and with the same
correlations functions.

In the same way we can expect  that  $Q'_s$  
corresponds to the deformed
$BCFT''$ with the boundary interaction term
\begin{equation}\label{b2}
S_{bound}=\lambda \int d\tau
 \cos \left(
\frac{2}{\sqrt{\alpha'}}X^1(\tau)\right)
\ . 
\end{equation}
This BCFT was extensively studied in 
\cite{Callan:1993mw,Callan:1994ub,Polchinski:my,
Recknagel:1998ih,Gaberdiel:2001zq}.

 It is remarkable that
 (\ref{Proll})
can be expressed in 
the pure gauge form which could lead to the 
suspicion that this solution does not give
any new physical information. We would like to
argue that this is not the case.
As is well known   the  string field theory action
(\ref{actionW})  is invariant 
under the small gauge transformations 
\begin{equation}
\delta \Phi=Q\Lambda-\Lambda \star \Phi+
\Phi\star \Lambda \ ,
\end{equation}
where $\Lambda $ is ghost number zero string field.
 On the other hand this action is not generally
invariant under the large gauge transformations 
\begin{equation}
\Phi'=e^{-\Lambda}\star Q(e^{\Lambda})+
e^{-\Lambda}
\star \Phi\star e^{\Lambda} \ .
\end{equation}
As is well known there is a sharp distinction between
the small gauge transformations and the large ones, 
for very nice discussion, 
see  \cite{Harvey:1996ur}. As was argued there,
the small gauge transformation describes redundancy in
our description of the theory. 
On the other hand, large
gauge transformations are true symmetries that relate
different solutions in given gauge theory which in our
case is the open bosonic  string field theory. 
\section{Spectrum of the fluctuations
around the new background}
In the previous section we have found an exact
solution of the string field theory and
after expansion above this solution 
an action for fluctuation modes that has
the same form as the original one
however with new,   shifted BRST operator
$Q'$. The
form of this operator suggests that the
string field theory formulated around this
classical solution should be equivalent to the
string field theory formulated around
the new $BCFT''$ that arises from the
original  $BCFT$ by exactly marginal deformation.
In this section we will try to give arguments
that could support this claim. More precisely,
we will try to determine  spectrum of 
 fluctuation modes around the classical
solution and these modes should correspond to 
the states in the deformed $BCFT''$.  

We start with the well known fact that
  fluctuation modes around the classical
solution obey linearised SFT equation of motion
\begin{equation}
Q'\Psi=Q(\Psi)-\mathcal{D}(\Psi)=0 \ .
\end{equation}
To proceed observe that the new BRST operator $Q'$
can be written as
\begin{equation}
Q'(X)=e^{K}(Q(e^{-K}(X)) \ .
\end{equation}
We see that it is natural to propose 
following form of the fluctuation field 
\begin{equation}\label{flu}
\Psi=e^{K}(\Phi)  \
\end{equation}
since then we get
\begin{equation}
Q'(\Psi)=0=e^{K}Q(e^{-K}e^{K}(\Phi))=0
\Rightarrow 
Q(\Phi)=0 \ .
\end{equation}
We will see that the consistency 
of the definition of the action of $K$ on
the state $\Psi$ and the requirement of
the well defined vertex operator 
corresponding to the string field (\ref{flu})
  will lead to the
constraints on the allowed spectrum of the 
fluctuation field.

In this section we restrict ourselves to
operator $K_{2s}\equiv K$. The extension to
time dependent solution would be straightforward,
following \cite{Sen:2002vv}.
So that let us consider following
operator 
\begin{equation}
K=\frac{1}{2\pi i}
\int_C dz\left(:\exp \left(i\frac{2}{\sqrt{\alpha'}}
X^1_L(z)\right):+:
\exp \left(-i\frac{2}{\sqrt{\alpha'}}
X^1_L(z)\right):\right)\equiv K_++K_-
 \ .
\end{equation}
Let us consider the ground state vertex
operator of original  $BCFT$
\footnote{In this section we will write
explicitly the symbol of normal ordering
$:\ : $.}
\begin{eqnarray}\label{phion}
\Phi(0,0)=c(0):e^{ik_{\mu}X^{\mu}(0,0)}:
=c(0):e^{i2k_{\mu}X^{\mu}_L(0)}: \ , \nonumber \\
Q(\Phi(0,0))=0 
\Rightarrow -k_0^2+k_1^2=\frac{1}{\alpha'} \ ,
\nonumber \\
\end{eqnarray}
where we consider the dependence on $k_1$ only. 
In the following we will use these formulas
\begin{eqnarray}\label{OPE2k}
:\exp \left(i\frac{2}{\sqrt{\alpha'}}
X^1_L(z)\right)::
e^{2ik_{\mu}X^{\mu}_L(0)}:=z^{2k\sqrt{\alpha'}}
:\exp \left(i\frac{2}{\sqrt{\alpha'}}
X^1_L(z)\right)
e^{2ik_{\mu}X^{\mu}_L(0)}: \ , \nonumber \\
:\exp \left(-i\frac{2}{\sqrt{\alpha'}}
X^1_L(z)\right)::
e^{2ik_{\mu}X^{\mu}_L(0)}:=z^{-2k\sqrt{\alpha'}}
:\exp \left(-i\frac{2}{\sqrt{\alpha'}}
X^1_L(z)\right)
e^{2ik_{\mu}X^{\mu}_L(0)}: \ . \nonumber \\
\end{eqnarray}
We use these OPE in the calculation of
the commutators $[K,\Phi(0,0)]$ that arise
from the formula
\begin{equation}
\Psi=e^{K}(\Phi)=e^{K}(\Phi(0,0)\ket{0})=
e^{K}\Phi(0,0)e^{-K}e^{K}\ket{0}=
e^{K}\Phi(0,0)e^{-K}\ket{0} \ .
\end{equation}
 Now we have
\begin{equation}
\Psi(0,0)=e^{K}\Phi(0,0)e^{-K}=
\Phi(0,0)+\sum_{N=1}^\infty
\frac{1}{N!}
\overbrace{[K,[\dots ,[K,\Phi(0,0)]]]}^N\ .
\end{equation}
The commutator $[K,\Phi]$ is given by contour
integral around $z=0$. To perform this calculation
we use OPE
(\ref{OPE2k}). However in  order  these integrals
to be well defined we see that the power of $z$ should
be integer. This requirement together with (\ref{phion}) 
gives quantisation of the momentum $k_1\equiv k$
\begin{equation}
k=\frac{n}{2\sqrt{\alpha'}} \ , 
n=\pm 2,\pm 3, \pm 4 \dots \ .
\end{equation}
We can interpret this result as a
 selection rule for  momentum states.
In fact the same situation arises in the study of
the BCFT with the boundary integration given
(\ref{b2}). 

Let us start with the first allowed 
state  with $n=2$
\begin{equation}
\Phi_2(0,0)=:e^{\frac{2i}{\sqrt{\alpha'}}X_L(0)}: \ .
\end{equation} 
For this state we get
\begin{eqnarray}
[K,\Phi_2(0,0)]=[K_-,\Phi_2(0,0)]=
\frac{1}{2\pi i}\oint
dw \frac{1}{w^2}:e^{\frac{-2i}{\sqrt{\alpha'}}
X_L(w)}e^{\frac{2i}{\sqrt{\alpha'}}X_L(0)}:=
\nonumber \\
=\frac{\partial}{\partial z}
\left[\frac{1}{2\pi i}\oint
dw \frac{1}{w-z}
:e^{-\frac{2i}{\sqrt{\alpha'}}
X_L(w)}\right]_{z=0}e^{\frac{2i}{\sqrt{\alpha'}}X_L(0)}:
=\nonumber \\
=-\frac{2i}{\sqrt{\alpha'}}\partial_w X_L(0)
:e^{-\frac{2i}{\sqrt{\alpha'}}
X_L(0)}e^{\frac{2i}{\sqrt{\alpha'}}X_L(0)}:
=-\frac{2i}{\sqrt{\alpha'}}\partial_w X_L(0)
\nonumber \\
\end{eqnarray}
and also
\begin{eqnarray}
[K,[K,\Phi_2]]=
[K,-\frac{2i}{\sqrt{\alpha'}}\partial_w X_L(0)
]=\nonumber \\
=-\frac{2i}{\sqrt{\alpha'}}
\frac{1}{2\pi i}\oint dw
\left(\frac{i\sqrt{\alpha'}}{w}
e^{\frac{2i}{\sqrt{\alpha'}}X_L(w)}-
\frac{i\sqrt{\alpha'}}{w}
e^{-\frac{2i}{\sqrt{\alpha'}}X_L(w)}\right)=
\nonumber \\
=2\left(e^{\frac{2i}{\sqrt{\alpha'}}X_L(0)}-
e^{-\frac{2i}{\sqrt{\alpha'}}X_L(0)}
\right) \ , \nonumber \\
\left[K,[K,[K,\Phi_2(0,0)]]\right]=
-\frac{8i}{\sqrt{\alpha'}}\partial_w X_L(w) \ .
\nonumber \\
\end{eqnarray}
We see that further commutators will give
the same components as the three ones listen
above. In other words the fluctuation
field is given as 
\begin{eqnarray}
\Psi=\Phi_2+[K,\Phi_2]+\frac{1}{2}[K,[K,\Phi_2]]
+\frac{1}{6}[K,[K,[K,\Phi_2]]]+\dots=\nonumber \\
=e^{\frac{2i}{\sqrt{\alpha'}}X_L(0)}
-\frac{i}{\sqrt{\alpha'}}\partial_w X_L(0)+
e^{\frac{2i}{\sqrt{\alpha'}}X_L(0)}
-e^{\frac{-2i}{\sqrt{\alpha'}}X_L(0)}
-\frac{8i}{6\sqrt{\alpha'}}\partial_w X_L(0)
+\dots \ . \nonumber \\
\end{eqnarray}
We observe  that the fluctuation field $\Psi$ is 
given as linear combination of the modes that
belong to  the representation of the $SU(2)$ group with
the spin $j=1$. Exactly the same  spectrum of the states
arise in $BCFT''$  with the marginal interaction 
 (\ref{b2}).

Generally  we can argue as follows.
Let us consider operator  $\Phi_n(z)=e^{\frac{in}{\sqrt{\alpha'}}
X_L(z)}$ with the spin equal to $\frac{n}{2}$ as
can be seen from
\begin{equation}
\left[J_3,\Phi_n(0)\right]=
\frac{i}{\sqrt{\alpha'}}\oint
\frac{dw}{2\pi i}
\partial_w X_L(w)e^{\frac{in}{\sqrt{\alpha'}}
X_L(z)}=\frac{n}{2}\oint \frac{dw}{2\pi i}
\frac{1}{w-z}e^{\frac{in}{\sqrt{\alpha'}}X_L(z)}=
\frac{n}{2}
\Phi_n(0,0) \ ,
\end{equation}
where
\begin{equation}
 J_3=
\frac{i}{\sqrt{\alpha'}}\int
dw \partial X^1_L(w)
 \ .
\end{equation}
From (\ref{OPE2k}) we get that $\Phi_n$ is 
 annihilated  by $K_+$ for $n\geq 0$ and also following
commutation relations
\begin{equation}
[K_+,K_-]=2J_3 \ ,
[J_3,K_+]=K_+ \ , [J_3,K_-]=-K_-  \ .
\end{equation}
Then  $[K,\Phi_n]=[K_-,\Phi_n]$ and  the resulting operator
will have the spin equal to $\frac{n}{2}-1$ as
follows from
\begin{eqnarray}
\left[J_3,[K_-,\Phi_n]\right]+
\left[K_-,[\Phi_n,J_3]\right]+
\left[\Phi_n,[J_3,K_-]\right]=0 
    \Rightarrow \nonumber \\
\Rightarrow
\left[J_3,[K_-,\Phi_n]\right]=\left(\frac{n}{2}
-1\right)[K_-,\Phi_n] \ . \nonumber \\
\end{eqnarray}
Then the successive action of $K$ on
$\Phi_n$ will give  states with
lower spin and hence the 
fluctuation field is linear combination of
states that form  an  irreducible representation of
$SU(2)$ gauge group with the spin $n/2$.
The same spectrum of states arises
in the $BCFT''$ with the marginal interaction
on the boundary (\ref{b2}). 
We think that this coincidence supports
our conjecture  based on the form of the shifted
BRST operator that the string field theory
for fluctuation modes around the classical
solution is related to he string field
theory defined on the background $BCFT''$
that arises from the original one
by marginal interaction inserted on the real line.
Note that we do not claim  that the string field
action for fluctuation modes  around the classical solution is
exactly the same as
the string field theory action formulated around the
new background BCFT. To find relation between
these two actions  we should perform
the same profound analysis as in 
\cite{Sen:1994kx,Sen:1993mh,Sen:1990na,Sen:1990hh,Sen:1992pw}. 
We return to this question in the future.

\section{Conclusion}\label{fourth}
In this short note we have proposed an  exact solution
of the open bosonic string field theory. 
The basic idea of our approach is
to find such an operator of conformal weight one that
has nontrivial commutation relation with the BRST operator.
The condition that this operator has conformal dimension
equal to one is important since only in this case this
operator obeys (\ref{derK}), (\ref{derKI}) and 
then we can easily find the form of the shifted BRST operator.
The form of  shifted BRST operator suggests
that  SFT for fluctuation field above the classical
solution is related to the
SFT defined on  $BCFT''$ that arises from the
original $BCFT$ by insertion of the marginal 
deformation on the boundary of the world-sheet. 
In order to  support this claim we have performed an
 analysis of
the linearised equation of motion of the SFT around
the classical solution and we have argued that
the spectrum of the fluctuation modes agree
with the spectrum of states in $BCFT''$.

Of course, as we have also stressed in many places
in this paper, we cannot claim that we have really
prove that the SFT action for fluctuation modes is
exactly the same as the SFT action  formulated around
background $BCFT''$. This  can be easily seen from
the fact that the fluctuation states and correlation
functions in the shifted action 
are still formulated using the original
$BCFT$. 
In order to find exact relation between
action for fluctuation modes and SFT action
defined on $BCFT''$ we 
should perform the same careful analysis
as  in  \cite{Sen:1994kx,Sen:1993mh,Sen:1990na,Sen:1990hh,Sen:1992pw}.  
This problem is currently under investigation
and we  hope to report our result in near future.

We must also stress that there is important
limitation of our solution when we consider it
in the context rolling tachyon 
\cite{Sen:2002nu,Sen:2002in,Sen:2002an,
Sen:2002vv,Mukhopadhyay:2002en}.
Since this is solution of the SFT 
in the limit $g_s\rightarrow 0$  we
do not take into account the 
coupling between closed a open strings in 
the process of the rolling tachyon. When this coupling
is considered  the situation is
much more involved and interesting, as was shown in 
recent papers
\cite{Okuda:2002yd,Strominger:2002pc,Sen:2002qa}. 
It would be
very nice to extend our analysis to the case of open-closed 
string field theory 
\cite{Zwiebach:1997fe}.
It would be also interesting to search for an exact solution in
 the string field theory that would lead to the 
more general boundary perturbation as was studied in
\cite{Sen:2002vv}.

To conclude, we
 hope that our modest contribution 
presented in this paper 
could be helpful in solving these intriguing problems. 

{\bf Acknowledgment}
I would like to thank Ulf Danielsson and Ulf Lindstr\"om
for their support in my research.
This work is partly supported 
by EU contract HPRN-CT-2000-00122.
    
    \end{document}